\documentclass[]{article}
\usepackage{multicol}
\usepackage{apj}

\def\listitem{\par \hangindent=50pt\hangafter=1
     $\ $\hbox to 20pt{\hfil $\bullet$ \hfil}}

\def\puncspace{\ifmmode\,\else{\ifcat.\C{\if.\C\else\if,\C\else\if?\C\else%
\if:\C\else\if;\C\else\if-\C\else\if)\C\else\if/\C\else\if]\C\else\if'\C%
\else\space\fi\fi\fi\fi\fi\fi\fi\fi\fi\fi}%
\else\if\empty\C\else\if\space\C\else\space\fi\fi\fi}\fi}
\def\SP{\let\\=\empty\futurelet\C\puncspace}

\def\degree{$^\circ$\SP}
\def\h1{$h^{-1}$\SP}

\def\apjs{ApJS,}

\def\lsim{~\rlap{$<$}{\lower 1.0ex\hbox{$\sim$}}}
\def\gsim{~\rlap{$>$}{\lower 1.0ex\hbox{$\sim$}}}
\def\void#1{{}}

\parskip=\baselineskip
\newcommand{\mincir}{\raise -2.truept\hbox{\rlap{\hbox{$\sim$}}\raise5.truept
\hbox{$<$}\ }}
\newcommand{\magcir}{\raise -2.truept\hbox{\rlap{\hbox{$\sim$}}\raise5.truept
\hbox{$>$}\ }}
\newcommand{\minmag}{\raise-2.truept\hbox{\rlap{\hbox{$<$}}\raise 6.truept\hbox
{$>$}\ }}
\def\be{\begin{equation}}
\def\ee{\end{equation}}

\newcommand{\ba}{\begin{eqnarray}}
\newcommand{\ea}{\end{eqnarray}}
\newcommand{\brr}{\begin{array}}
\newcommand{\err}{\end{array}}
\newcommand{\bc}{\begin{center}}
\newcommand{\ec}{\end{center}}
\newcommand{\br}{\mbox{\bf r}}

\newcommand{\lb}{{\left<\right.}}
\newcommand{\rb}{{\left.\right>}}
\newcommand{\vel}{\rm \,km\,s^{-1}}

\def\hm{$h^{-1}$Mpc\SP}

\def\listitem{\par \hangindent=50pt\hangafter=1
     $\ $\hbox to 20pt{\hfil $\ubfllet$ \hfil}}
\def\puncspace{\ifmmode\,\else{\ifcat.\C{\if.\C\else\if,\C\else\if?\C\else%
\if:\C\else\if;\C\else\if-\C\else\if)\C\else\if/\C\else\if]\C\else\if'\C%
\else\space\fi\fi\fi\fi\fi\fi\fi\fi\fi\fi}%
\else\if\empty\C\else\if\space\C\else\space\fi\fi\fi}\fi}
\def\SP{\let\\=\empty\futurelet\C\puncspace}

\def\degree{$^\circ$\SP}
\def\h1{$h^{-1}$\SP}

\begin{document}
\vspace{15mm}                                                                  
\begin{center}
\uppercase{\large{\bf Correlation Analysis of SFI Peculiar Velocities}}\\
\vspace*{1.5ex}
{\sc Stefano Borgani} \\
{\small 
INFN, Sezione di Trieste, c/o Dipartimento di Astronomia
dell'Universit\`a, via Tiepolo 11,
I-34100 Trieste, Italy\\
INFN, Sezione di Perugia, c/o Dipartimento di
Fisica dell'Universit\`a, via A. Pascoli, I-06123 Perugia, Italy\\
Electronic mail: borgani@ts.astro.it}\\
\vspace*{1.5ex}
{\sc Luiz N. da Costa} \\
{\small 
European Southern Observatory, Karl
Schwarzschild Str. 2, D--85748 Garching b. M\"unchen, Germany\\
Observat\'orio Nacional, Rua Gen. J. Cristino 77,  S\~ao Cristov\~ao, Rio
de Janeiro, Brazil\\ Electronic mail: ldacosta@eso.org}\\
\vspace*{1.5ex}
{\sc Idit Zehavi}\\
{\small NASA/Fermilab Astrophysics Group, Fermi National Accelerator 
Laboratory, \\Box 500, Batavia, IL 60510-0500, U.S.A.\\ 
Electronic mail: iditz@simone.fnal.gov}\\
\vspace*{1.5ex}
{\sc Riccardo Giovanelli and Martha P. Haynes}\\
{\small Center for Radiophysics and Space Research
and National Astronomy and Ionosphere Center\footnote{The National
Astronomy and Ionosphere Center is operated by Cornell University
under a cooperative agreement with the National Science Foundation.},\\
Cornell University, Ithaca, NY 14953\\Electronic mail:
riccardo@astrosun.tn.cornell.edu, haynes@astrosun.tn.cornell.edu}\\
\vspace*{1.5ex}
{\sc Wolfram Freudling}\\
{\small Space Telescope--European Coordinating Facility, 
European Southern Observatory, \\Karl Schwarzschild Str. 2, D--85748
Garching b. M\"unchen, Germany\\Electronic mail: wfreudli@eso.org}\\
\vspace*{1.5ex}
{\sc Gary Wegner}\\
{\small Department of Physics and Astronomy, Dartmouth College, Hanover, 
NH 03755, U.S.A.//Electronic mail: wegner@kayz.dartmouth.edu}\\
\vspace*{1.5ex}
{\sc John J. Salzer}\\
{\small Department of Astronomy, Wesleyan University, Middletown, CT
06459, U.S.A.\\Electronic mail: slaz@parcha.astro.wesleyan.edu}\\
\vspace*{1.ex} 
\end{center}
\vspace*{-6pt}
 
\begin{abstract}
We present results of a statistical analysis of the SFI catalog of
peculiar velocities, a recently completed survey of spiral field
galaxies with I-band Tully-Fisher distances.  The velocity field
statistic utilized is the velocity correlation function, $\psi_1(r)$,
originally introduced by G\'orski et al. (1989).  The analysis is
performed in redshift space, so as to circumvent potential ambiguities
connected with inhomogeneous Malmquist bias corrections.  The results
from the SFI sample are compared with linear--theory predictions for a
class of cosmological models. We generate a large set of mock samples,
extracted from N--body simulations, which are used to assess the
reliability of our analysis and to estimate the associated
uncertainties.  We assume a class of CDM--like power spectrum models,
specified by $\sigma_8$, the r.m.s. fluctuation amplitude within a
sphere of $8$\hm radius, and by the shape parameter $\Gamma$.
Defining $\eta_8=\sigma_8\Omega_0^{0.6}$, we find that the measured
$\psi_1(r)$ implies a degenerate constraint in the $\eta_8$--$\Gamma$
plane, with $\eta_8=0.3 \pm 0.1 (\Gamma/0.2)^{0.5}$, at the $2\sigma$
level, for the inverse Tully--Fisher (ITF) calibration presented in
this paper. We investigate by how much this constraint changes as we
account for uncertainties in the analysis method and uncertainties in
the distance indicator, and consider alternative ITF calibrations.  We
find that both changing the error weighting scheme and selecting
galaxies according to different limiting line--widths has a negligible
effect. On the contrary, the model constraints are quite sensitive to
the ITF calibration.  The other ITF calibrations by Giovanelli et
al. (1997) and da Costa et al.  (1998) both give, for $\Gamma=0.2$,
$\eta_8\simeq 0.6$ as the best--fitting value.
\vspace*{6pt}
\noindent
FERMILAB-Pub-99/133-A
\end{abstract}

\begin{multicols}{2} 

\section{INTRODUCTION}
\label{intro}

The peculiar velocity field of galaxies provides a very powerful way
of probing mass fluctuations on intermediate to large scales ($\lsim$
100~\h1 Mpc, $h$ being the Hubble constant in units of 100$\vel$
Mpc$^{-1}$), as it is sensitive primarily to large scale density
fluctuations. Therefore, studies of cosmic flows can be used to
constrain the amplitude of the large--scale mass power-spectrum, thus
complementing the information on intermediate scales, between those
probed by redshift surveys and those sampled by anisotropies in the
cosmic microwave background (CMB) as observed by COBE (see the review
by Dekel 1994).  Another advantage in studying the velocity field is
that it is measured on scales where linear approximation to
gravitational instability is expected to hold, thus allowing one to
explore more thoroughly the parameter space of cosmological models. We
can parameterize the fluctuation power spectrum in terms of the
r.m.s. fluctuation within spheres of $8$\hm, $\sigma_8$, and of a
shape parameter $\Gamma$. Then, according to linear theory, the
typical amplitude of the peculiar velocity on a given scale is
proportional to $\eta_8 \,f(\Gamma,R)$, where
$\eta_8=\sigma_8\Omega_m^{0.6}$ (following the notation of Chiu,
Ostriker \& Strauss 1998; $\Omega_m$ here is the matter density
parameter) and $f(\Gamma,R)$ is a quantity which depends on the power
spectrum shape and on the scale $R$ at which the velocity field is
probed.

Several statistical characterizations of the peculiar velocity fields
have been proposed in the last decade, with the aim of providing more
robust constraints on cosmological scenarios, as newer and larger data
sets came to completion (e.g., Strauss \& Willick 1995, for a
review). Among such statistical measures, in this paper we will
concentrate on the velocity correlation function, which has been
introduced for turbulence studies by Monin \& Yaglom (1975) and
borrowed for cosmology by Peebles (1980; cf. also G\'orski 1988).  We
will apply this statistic to the SFI sample, a recently completed
homogeneous all--sky survey of Sbc-Sc galaxies with I--band
Tully--Fisher (TF) distances (Giovanelli et al. 1997a; Haynes et
al. 1999a,b, H99a,b).

A first application of the velocity correlation statistics to
observational data was realized by G\'orski et al. (1989, G89
hereafter; see also Groth, Juszkiewicz \& Ostriker 1989), who analyzed
the spiral galaxy sample by Aaronson, Huchra \& Mould (1979) and the
elliptical galaxy sample by Burstein et al. (1987), finding
substantial discrepancies between the results obtained from these two
data sets. Tormen et al.  (1993, T93) analyzed the correlation
statistics of the Mark II sample, with results favoring $\eta_8\simeq
0.7$ for scale--invariant CDM models. Kolatt \& Dekel (1996) estimated
the matter power--spectrum implied by the POTENT reconstruction of the
Mark III data (Willick et al. 1997) and found $\eta_8\simeq 0.7$--0.8.
More recently, maximum--likelihood analyses, estimating the mass
power-spectrum that gives rise to the observed peculiar velocities,
have been performed by Zaroubi et al. (1997) on the Mark III sample
and by Freudling et al. (1999, FZ99) on the SFI sample. Both analyses
consistently find $\eta_8 \simeq 0.8 \pm 0.2$ ($90\%$ c.l.), quite
independent of the power--spectrum shape. These results point toward
high--amplitude fluctuations, thus somewhat at variance with results
from the r.m.s. cluster peculiar velocity (e.g., Borgani et al. 1997;
Watkins 1997) and with constraints from the local cluster abundance
(e.g., Eke, Cole \& Frenk 1996; Girardi et al. 1998), which indicate
lower values.

Studies of the peculiar velocity can also be combined with analyses of
all--sky redshift surveys to investigate the relation between the
galaxy and underlying mass distributions, a key ingredient for
understanding galaxy biasing. Comparisons between the measured
peculiar velocities or the recovered densities with those predicted
from all--sky redshift surveys are commonly used to estimate the
parameter $\beta = \Omega_m^{0.6}/b$, under the assumption of linear
biasing with a bias factor $b$.  Several estimates of $\beta$ have
been presented in the literature (e.g., da Costa et al. 1998; Willick
\& Strauss 1998; Branchini et al. 1999, and references therein) based
on comparisons between the velocity fields directly inferred from TF
data and recovered from galaxy density field in the {\sl IRAS} 1.2 Jy
(Fisher et al. 1995) and PSCz survey. Such analyses generally find
$\beta$ values in the range 0.5--0.7. Taking
$b=\sigma_{8,IRAS}/\sigma_8$, these results would imply $\eta_8\simeq
0.35$--0.50 for $\sigma_{8,IRAS}\simeq 0.7$ (Fisher et al. 1994). On
the other hand, analyses based on the comparison of density fields
provide values of $\beta$ as large as 0.9 (e.g., Sigad et al. 1998).
The interpretation of the $\beta$ values is further complicated if
galaxy biasing is better described by a stochastic, nonlinear process
(e.g., Dekel \& Lahav 1999).

The aim of this paper is to perform a detailed analysis of the
velocity correlation function for the SFI sample and to derive the
resulting constraints on large--scale structure formation models. The
comparison to theoretical expectations is based on linear--theory
predictions and we resort to large--scale N--body simulations to
verify the reliability of our analysis and to estimate the associated
errors, contributed by both the cosmic variance and by the scatter in
the TF relation.

In our analysis, we choose to use redshift--space information as the
indicator of distance for the SFI galaxies, so as to avoid the
associated Malmquist bias arising from the intrinsic scatter of the
distance indicator when using the inferred distances (cf. Freudling et
al. 1995, for a discussion on bias corrections in the SFI sample). The
forward TF relation, obtained by regressing the apparent magnitudes
over the line--width, in this case, is still susceptible to selection
bias due to the imposed magnitude--limit. Using the inverse relation,
i.e. fitting the line--width as a function of the apparent magnitude,
avoids this selection bias, as long as the sample selection is
independent of the line--width (see \S6 of Strauss \& Willick 1995,
and references therein).  For this reason, we perform our analysis in
redshift--space by using peculiar velocities estimated from the
inverse Tully--Fisher (ITF) relation.

The outline of the paper is as follows. In Section 2 we provide a
basic description of the SFI sample and present the ITF calibrations
on which our analysis is based. Section 3 contains a brief
introduction to the velocity correlation formalism and presents the
results of its application to the SFI data. In Section 4 we present
the velocity correlation analysis of our mock samples.  In Section 5
we derive the resulting constraints on cosmological models and discuss
the impact of systematic effects in both the sample definition and the
correlation analysis method. We summarize our main conclusions in
Section 6.

\section{THE SFI SAMPLE}
\label{data}

The TF data defining the sample used here consists of two main
datasets: a subset of the Mathewson, Ford \& Buchhorn (1992) survey
with about 1200 galaxies with I-band photometry and measured
rotational velocities, either from radio observations of 21-cm
line--widths or optical rotation curves; the SFI I--band TF
redshift--distance survey of about 1300 Sbc-Sc field galaxies. The SFI
sample consists of galaxies with inclination $\gsim$ 45\degree north
of $\delta<-45^\circ$ and galactic latitudes $|b| > 10^{\circ}$.  The
original Mathewson et al. (1992) measurements of magnitude and
rotational velocities were converted into the SFI system using about
200 to 300 common galaxies.

In addition to the field galaxies, roughly 800 galaxies covering a
broader range of morphological types were observed in the field of 24
clusters (Giovanelli et al. 1997a,b; SCI sample). After careful
membership assignment, cluster galaxies were used to derive a combined
TF relation corrected for Malmquist bias and bias introduced by
incompleteness and different morphological mix.  In order to perform
our analysis in redshift--space, we consider the inverse TF relation
(ITF, hereafter) between the absolute magnitude $M$ and the full
line--width $W$,
\be 
M\, =\, a+b(\log W - 2.5)\,,
\label{eq:itf}
\ee with $a=-20.95$ and $b=-7.94$ (here $W$ is expressed in units of
km $s^{-1}$ and we assume a Hubble constant of 100 km $s^{-1}$
Mpc$^{-1}$).  This relation has the same slope as that originally
provided by Giovanelli et al. (1997b, G97 hereafter), whose 
zero--point, $a=-21.10$, is 0.15 magnitudes smaller. This
difference is due to a new determination of the velocity widths and to
the removal of 71 galaxies due to poor photometry, poor line--widths
or obvious misidentification (cf.  H99a,b). The $1\sigma$ uncertainty
in the zero--point has been estimated by G97 to be about 0.05
magnitudes, when combining statistical uncertainties in the TF fitting
and uncertainties in defining the cluster reference frame with a
finite number (24) of such objects.  This uncertainty does not however
include possible systematics associated with the processing of the raw
data or with difference between the TF relation of clusters and field
galaxies, or potential deviations of our local universe from a global
Hubble flow (e.g. Zehavi et al. 1998, but see also Giovanelli et
al. 1999).

We note that careful analysis of the TF relation for galaxies in
clusters suggests that the scatter depends on the line--width.  This
dependence is modeled by letting the error in the estimated distance
$r_i$ of the $i$--th galaxy to be $\epsilon_i = \Delta(W_i) r_i$,
where $\Delta(W_i)$ is the fractional error in the distance as
estimated from the scatter about the ITF relation as a function of the
measured line--width of the galaxy (G97, cf. also Willick et al. 1997
and Willick \& Strauss 1998).  The resulting errors are estimated to
be in the range 15--20\%.

Unless otherwise specified and following da Costa et al. (1996) and
FZ99, we discard those ($\sim 7\%$) SFI galaxies with line--width
$\log W\le 2.25$, because of the limited reliability of the ITF
relation at such line--widths. We will also show the robustness of the
final results against changes in the assumed limiting line--width.
Furthermore, we restrict our analysis to the SFI subsample defined by
galaxies lying within $cz\le 6000\vel$. With such restrictions, the
final sample on which we base our analysis contains 974 galaxies.

A further alternative calibration of the ITF has been presented by da
Costa et al. (1998, dC98 hereafter), based on a comparison of the
velocity field of the SFI sample and that implied by the IRAS 1.2 Jy
survey. The resulting zero--point and slope of the ITF are $a=-21.11$
and $b=-8.55$, respectively. In the following, we will use the above
most recent ITF calibration as the reference one, but will show the
effect of taking the previous G97 and dC98 calibrations on the final
constraints on cosmological parameters.

\section{THE VELOCITY CORRELATION STATISTICS}
\label{cor}
The estimator for the velocity correlations that we will use in the
following is that introduced by G89 and is given by
\be
\psi_1(r)\,=\,{\sum_{|\br_i-\br_j|=r}w_iw_j u_iu_j\cos\vartheta_{ij} \over
  \sum_{|\br_1-\br_j|=r}w_iw_j \,\cos^2 \vartheta_{ij}}\,,
\label{eq:psi1}
\ee
where $\vartheta_{ij}$ is the angle between the direction of the
$i$-th and the $j$-th galaxy and the sums are over all the galaxy
pairs at separation $r$ in redshift space.  With the above definition,
the $\Psi_1(r)$ statistics is independent of any assumptions regarding
the velocity field, such as homogeneity and isotropy, and has been
shown by G89 to be rather robust to sampling fluctuations.  In
eq.(\ref{eq:psi1}) $u_i$ is the radial peculiar velocity of the
$i$--th galaxy and $w_i$ represents a suitable weight to be assigned
to it. The introduction of the weights is a slight modification of the
expression for $\psi_1$ provided by G89 (see also T93). Different
weighting schemes will be applied in the following: $(1)$ uniform
weighting, $w_i=1$; $(2)$ weighting galaxies according to their
distance--error, $w_i=1/\epsilon_i$; $(3)$ weighting according to
$w_i^2=1/(\epsilon_i^2+\sigma_f^2)$, where $\sigma_f^2$ is the
variance of the local velocity field. 

The quantity $\sigma_f$ can be
interpreted as a line--of--sight velocity dispersion and has
been introduced in order to model possible non--linearities, which
generates small--scale random motions within virialized regions. Such
motions, which would give rise to an uncorrelated velocity component,
are expected to be relatively unimportant for the SFI field galaxies,
whose peculiar velocity should not be much affected by virial
motions. A further possible interpretation of $\sigma_f$ is an
unrecognized distance--independent error, which is not accounted for
by the ITF scatter calibrated by using members of distant clusters
(e.g., Kaiser 1988).  FZ99 checked for such a term by having it as a
further degree of freedom to be constrained by a maximum likelihood
approach and found $\sigma_f=200\pm 120\vel$. When resorting to the
weighting scheme (3), we will take $\sigma_f=150\vel$, although our
final results are essentially insensitive to its choice.

As for the scheme $(1)$, its main drawback is that it assigns the same
weight to all objects, regardless of the uncertainty in the velocity
errors, which increase with distance. Although the methods $(2)$ and
$(3)$ overcome this limitation, they reduce the effective sampling
volume, and have been shown by Dekel, Bertschinger \& Faber (1990) to
overestimate the contribution of well sampled regions with respect to
under-sampled regions in the reconstruction of velocity fields. In the
following we will mainly base our analysis on the uniform--weighting
scheme, which is the least affected by cosmic scatter (see Section
\ref{test} below). 

As shown by G89, the ensemble average of $\psi_1(r)$ is given by
\be
\Psi_1(r)\,=\,\lb \psi_1(r)\rb\,=\,{\cal
A}(r)\Psi_\parallel(r)+\left[1-{\cal A}(r)\right]\Psi_\perp(r)\,,
\label{eq:ppsi1}
\ee
under the assumption of homogeneity and isotropy, 
where $\Psi_\parallel$ and $\Psi_\perp$ are the
radial and transverse correlation functions of the three--dimensional
peculiar velocity field. In linear theory, they are connected to the
power--spectrum of density fluctuations, $P(k)$, according to 
\ba
\Psi_\parallel(r) & = & {f(\Omega_m)^2\,H_0^2\over 2\pi^2}\,\int dk
\,P(k)\,\left[j_0(kr) - 2{j_1(kr)\over kr}\right ]\,; \nonumber \\
\Psi_\perp(r) & = & {f(\Omega_m)^2\,H_0^2\over 2\pi^2}\,\int dk
\,P(k)\,{j_1(kr)\over kr}\,,
\label{eq:psi}
\ea
where $j_i(x)$ is the $i$-th order spherical Bessel function and
$f(\Omega_m) \simeq \Omega_m^{0.6}$. 

The quantity $\cal A$ appearing in eq.(\ref{eq:ppsi1}) is
a moment of the selection function of the sample 
depending on the spatial distribution of galaxies 
according to 
\ba
& &\!\!\!\!\!\!\!\!\!\!\!\!\!\!\!\!{\cal A}(r)\,=\nonumber \\
& &\!\!\!\!\!\!\!\!\!\!\!\!\!\!\!\!{\sum_{|\br_i-\br_j|=r}w_i w_j\left[ r_i r_j(\cos
\vartheta_{ij}-1)+r^2\cos \vartheta_{ij}\right]\,\cos
\vartheta_{ij}\over
r^2 \sum_{|\br_i-\br_j|=r}w_i w_j\cos^2 \vartheta_{ij}}\,.
\label{eq:ar}
\ea
This quantity provides in a sense the relative contribution to
$\psi_1(r)$ from the radial and transverse components of the velocity
correlation.  The definition of eq.(\ref{eq:ar}) is slightly different
from that previously adopted by other authors, by including the galaxy
weights.

The advantage of using $\psi_1$ is that it can directly be calculated
from the observed radial velocities, without the need of any
additional assumption.  It can then be related to theory
(eq.[\ref{eq:ppsi1}],[\ref{eq:psi}]), taking into account the specific
sampling through eq.(\ref{eq:ar}). The geometrical factor ${\cal
A}(r)$ is plotted in Figure 1 for the three mentioned
weighting schemes. The net effect of a non--uniform weighting is that
of increasing ${\cal A}(r)$ at separations $\magcir 2000$ km s$^{-1}$.
This is the consequence of the fact that $\Psi_\perp$ takes relatively
more contribution than $\Psi_\parallel$ from large--scale fluctuations
(see, e.g., G\`orski 1988). Therefore, its contribution to $\Psi_1(r)$
is suppressed with the error weighting, which amounts to 
decreasing the effective volume of the sample.

\includegraphics{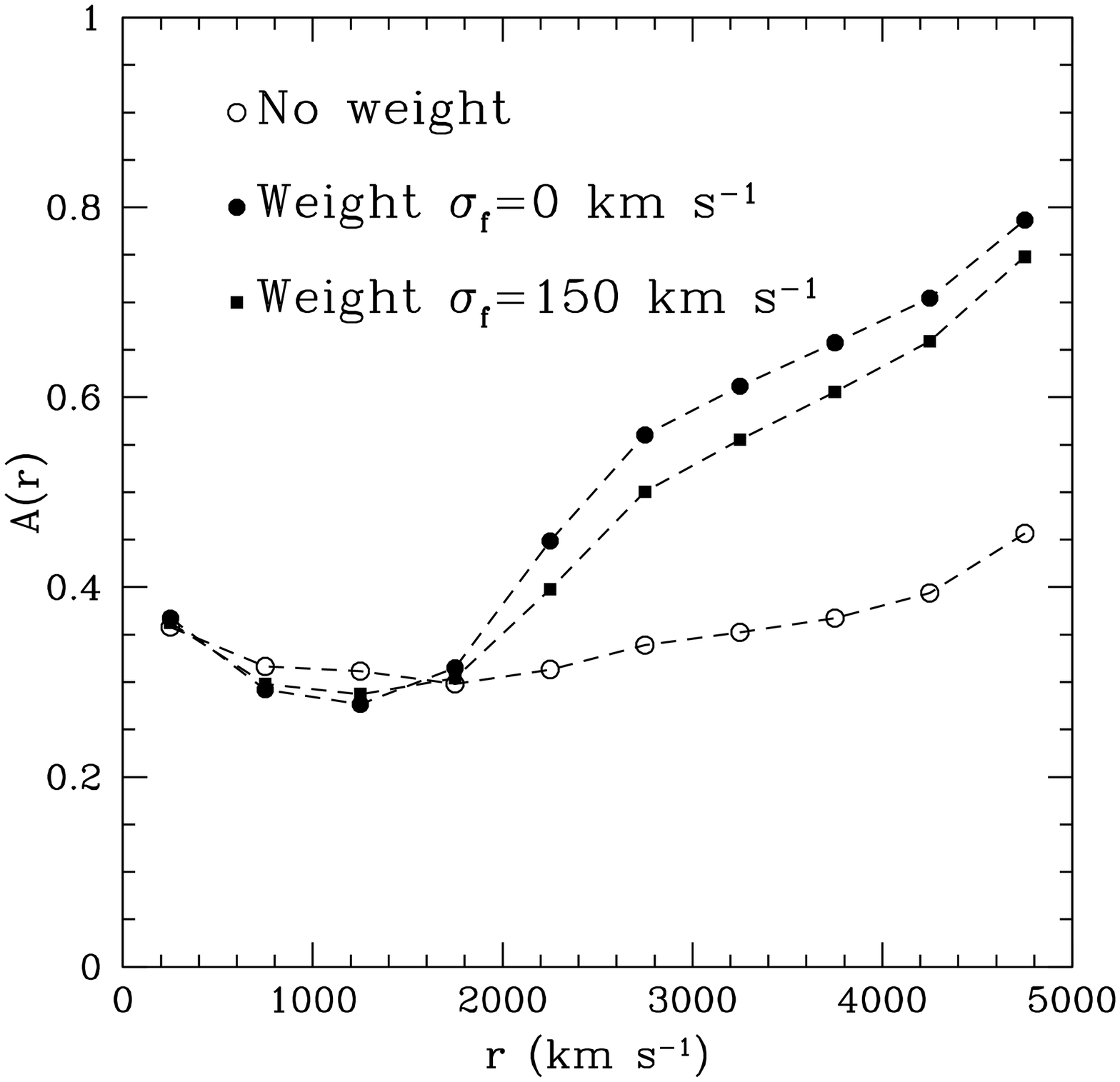}
$\ \ \ \ \ \ $\\
\vspace{9.4truecm}
$\ \ \ $\\
\vspace{-0.3truecm}

{\small\parindent=3.5mm {Fig.}~1.--- The geometrical factor ${\cal
A}(r)$ (eq.[\ref{eq:ar}]; see text), associated with the SFI sample,
for the three alternative weighting schemes.}
\vspace{5mm}

The velocity correlation function $\psi_1(r)$ for the SFI sample,
with the H99 calibration,
computed within bins of 500$\vel$, is plotted in Figure 2.
No errorbars are assigned here to $\psi_1(r)$. We
will discuss in the next section how to associate uncertainties to
model predictions, in order to provide confidence levels in the
estimate of cosmological parameters.  The upper panel shows the effect
of adopting different weighting schemes.  It is apparent that the
choice for $w_i$ has a marginal impact on the correlation signal.
This result might seem somewhat unexpected, in view of the different
$A(r)$ values for the weighted and unweighted cases.  However, these
differences appear only at rather large separations, $r\magcir
2000\vel$ (cf. Figure 1), where the value of $\psi_1$ for SFI rapidly
declines, thus making any difference among different weighting schemes
hardly detectable.  By comparing this result with that from the
real--space analysis of the Mark II sample by Tormen et al. (1993), it
turns out that the SFI sample produces a velocity correlation signal
which is at least a factor two smaller, although the corresponding
scales at which $\psi_1(r)$ approaches zero ($\simeq 3000$ km
s$^{-1}$) are similar.

\vspace{-0.4truecm}
\includegraphics{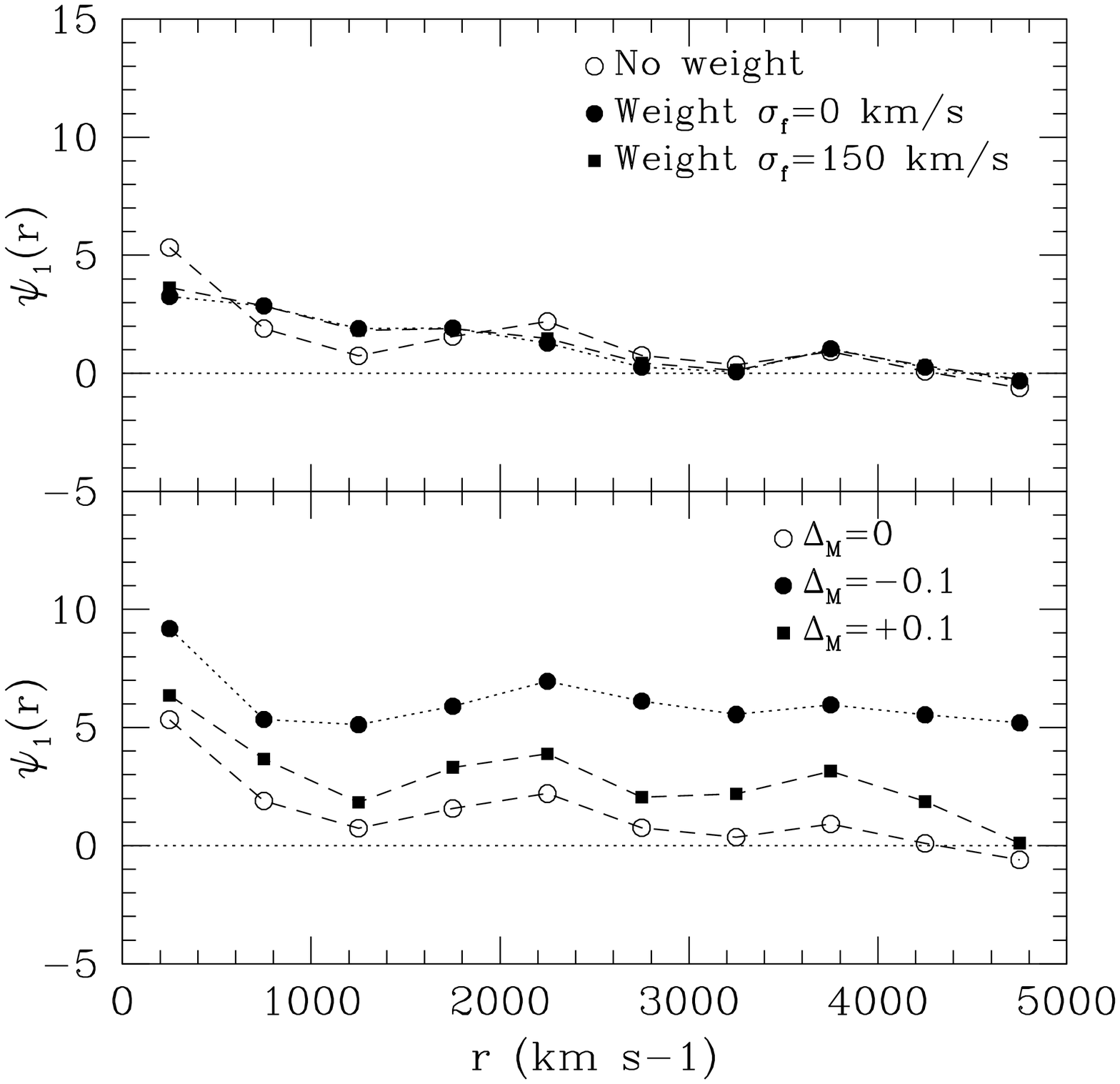}
$\ \ \ \ \ \ $\\
\vspace{9.4truecm}
$\ \ \ $\\
\vspace{-0.3truecm}

{\small\parindent=3.5mm {Fig.}~2.--- The velocity correlation
function, $\psi_1(r)$ (in units of $10^4\vel$), for the SFI
sample. The upper panel shows the effect of different galaxy weights,
while the lower panel shows the effect of changing by 0.1 magnitudes
the zero--point of the ITF relation, representing the $2\sigma$
uncertainty in its calibration (cf.  G97, H99a,b).  }
\vspace{5mm}

The lower panel of Fig. 2 shows the effect of changing the zero--point
of the ITF relation (eq.[\ref{eq:itf}]) by 0.1 magnitudes either way,
which corresponds to a change of $\epsilon\sim 2.5\%$ in the distances
or an additional global Hubble--like flow $\epsilon r$. This change
corresponds to the $2 \sigma$ formal statistical uncertainty estimated
from the analysis of the SCI sample of cluster galaxies (G97, H99a,b).
A global Hubble--like flow represents a coherent velocity field which
is characterized by a positive correlation (i.e., galaxies moving in
the same direction) on intermediate scales, $r\mincir 5000\vel$ and by
a negative correlation at the largest scales, $r\magcir 7000\vel$,
when the two galaxies of a pair are placed in the opposite
directions of the sample.

Alternative estimators of the velocity correlation statistics have
been applied by different authors. Groth et al. (1989; cf. also Kaiser
1988) considered the generic form for the velocity correlation tensor
under the assumption of homogeneous and isotropic velocity field,
$\Psi_{ij}(r)=\lb v_i(\vec x)v_j(\vec x-\vec
r)\rb=\Psi_\perp(r)\delta_{ij} +[\Psi_\parallel(r)-\Psi_\perp(r)]\hat
r_i\hat r_j$, where $\delta_{ij}$ is the Kronecker symbol. Then, they
obtained $\Psi_\perp$ and $\Psi_\parallel$ by a $\chi^2$--minimization
procedure to the data. G89 compared this method to their $\Psi_1(r)$
approach and showed that they produce comparable results, although the
former turns out to be noisier at large separations, $r\magcir
4000\vel$.

More recently, Ferreira et al. (1999) proposed a new method to
estimate the main galaxy pairwise velocity, $\vec v_{12}=\lb \vec
v(\vec x_1)-\vec v(\vec x_2)\rb$. This method, which has been so far
tested on N--body mock samples and is in the process of being applied
to real data sets, provides essentially constraints on
$\sigma_8^2\Omega_m^{0.6}$. Therefore, its combination with
linear--theory constraints on $\sigma_8\Omega_m^{0.6}$ could in
principle break the degeneracy between $\sigma_8$ and $\Omega_m$. Of
course, careful investigations are required in order to understand
whether available data are of sufficient quality and their systematics
and biases are enough under control to allow a reliable estimate of
$\sigma_8$ and $\Omega_m$ separately.

\section{ANALYSIS OF THE MOCK SAMPLES}
\label{test}
In order to explore extensively the model parameter space, we resort
in the following to linear theory as the means to compare model
predictions and SFI results. Two important issues need to be
addressed: $(a)$ the reliability of our analysis and specifically the
use of linear theory to predict the statistics of the velocity field,
and $(b)$ the estimate of the cosmic scatter and the observational
uncertainties associated with the SFI sampling, in order to establish
the confidence level for model exclusion.  For this purpose we use
large N--body simulations from which we extract sets of mock samples
which mimic the sampling and selection effects of the SFI sample.

\subsection {Generating the mock samples}
The parent N--body simulations from which we extract mock samples have
been run by using the publicly available adaptive $P^3M$ code by
Couchman (1991). We have run two simulations corresponding to two
different cosmological scenarios. The first model is a flat
low--density one with $\Omega_m=0.4$ ($\Lambda 0.4$).  The transfer
function used is that of Bardeen et al. (1986) [see eq.(\ref{eq:tk})
below], with the shape parameter $\Gamma$ set to $0.22$ and
$\sigma_8=0.87$.  The second model is an Einstein--de Sitter (EdS)
universe, with $\Gamma=0.43$, and $\sigma_8=1.2$. With the above
parameters, both models are consistent with the 4--year {\sl COBE}
normalization (e.g., Bunn \& White 1997), while the EdS model fails to
match the abundance of local galaxy clusters (e.g., Eke et al. 1996;
Girardi et al. 1998) and the shape of the galaxy power--spectrum
(e.g., Peacock \& Dodds 1994; Liddle et al. 1996).

Each simulation follows $128^3$ particles within a box of $250$\hm on
a side. The adopted Plummer softening scale, $\simeq 100\,h^{-1}$ kpc,
is more than adequate to describe the large--scale velocity field (see
Borgani et al. 1999, for a more detailed description of the
simulations). Velocity fields on scales of a few$\times 10$\hm, which
are of interest in this paper, receive a small but non--negligible
contribution from wavelengths larger than the adopted box
size. Furthermore, the volume of a single simulation can accommodate
only a rather small number of non--overlapping SFI mock samples (each
extending out to $cz=6000\vel$), so as to not allow a reliable
determination of cosmic variance.

In order to extend the dynamic range of our simulations to larger
scales, we resorted to the method proposed by Tormen \& Bertschinger
(1996) of adding longer waves to N--body outputs. This method, which
allows to generate non--periodic replicas of a parent box, is based on
the Zel'dovich approximation (Zel'dovich 1970) for computing the
contribution to particle displacements and velocities from waves
longer than the original box size. Cole (1997) showed that this
procedure is adequate to extend to larger scales the description of
peculiar velocities. In our analysis, we replicate the original box
three times along each spatial direction, which leads to a total of 27
replica and a final box of size $L=750$\hm, containing about
$5.7\times 10^7$ particles.

As a first step for mock sample extraction, we divide the large box
into $6^3$ smaller boxes of 125\hm on a side. At the center of each of
them we place an observer. After randomly choosing the orientation of
the ``galactic'' coordinate system, we select among the simulation
particles those which are closest to the position of real galaxies in
the SFI sample. In this way, we generate mock samples with the same
spatial distribution and number of galaxies as in the real SFI sample.
The ``true'' radial velocities in the mock samples are perturbed
according to the associated observational errors of the real catalog
and according to the assumed random velocity dispersion $\sigma_f$
(under the assumption that both contributions are independent Gaussian
variables). For each simulation, we generate two sets of mock samples,
based on assuming both $\sigma_f=0$ and $150\vel$. Since the final
results turn out to be essentially indistinguishable, we will present
for the mock sample analysis only results based on assuming a
vanishing $\sigma_f$.

We note that other authors (e.g., G89; Strauss, Cen \& Ostriker 1993;
T93) followed more sophisticated procedures to search for
``observers'' within simulations.  Such procedures involve selecting
observers so that local properties of the density and velocity field
resemble those observed for the Local Group of galaxies. However, T93
showed that applying such constraints on the observer selection does
not significantly alter the velocity correlation statistics for
realistic power spectra. Furthermore, the aim of our analysis is to
estimate how often the SFI correlation statistics can be observed in a
given cosmology assuming the variety of observers' characteristics to
be included into the cosmic variance which is appropriate for that
model.

\includegraphics{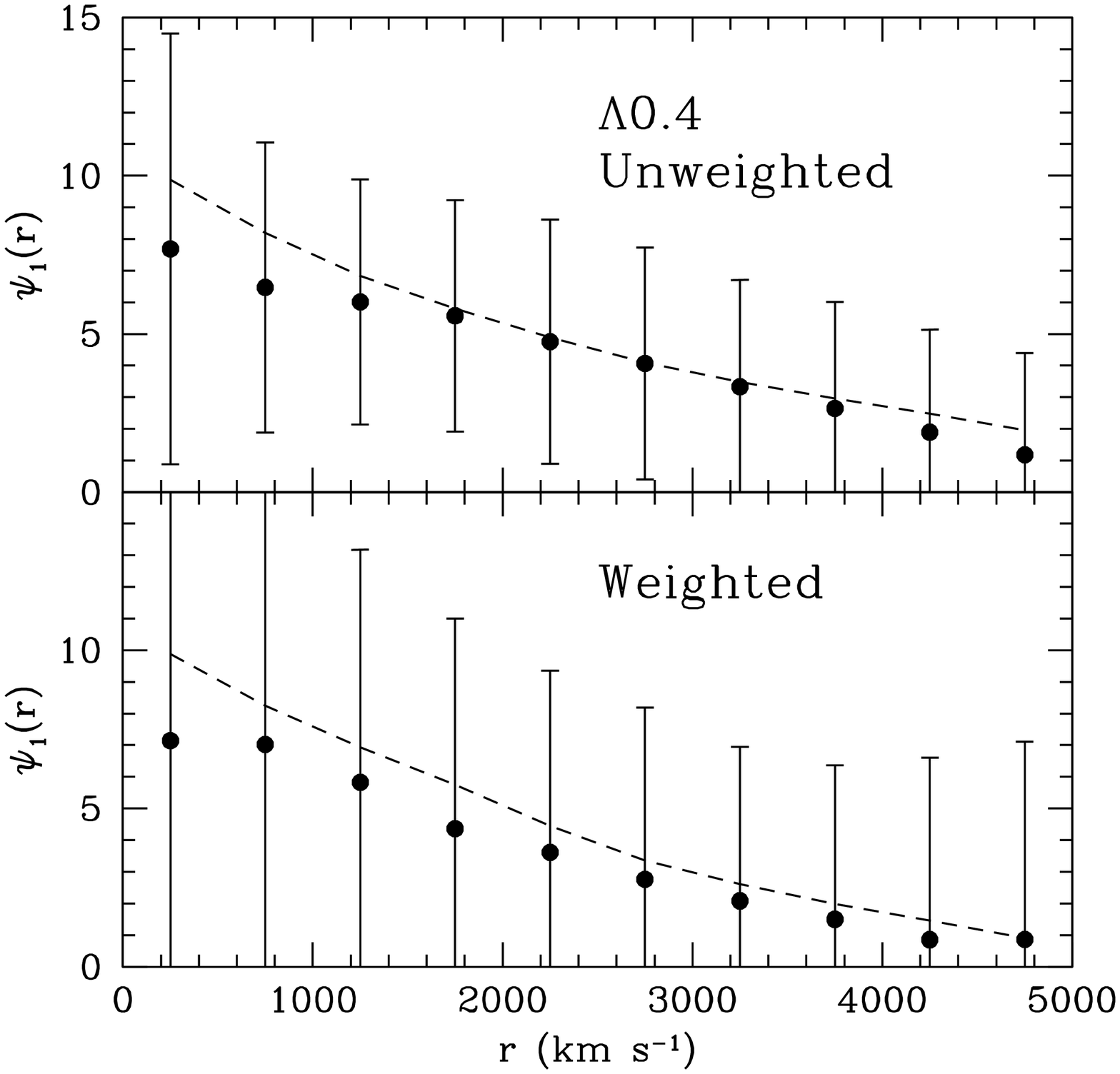}
$\ \ \ \ \ \ $\\
\vspace{8.4truecm}
$\ \ \ $\\
\vspace{-0.2truecm}

{\small\parindent=3.5mm {Fig.}~3.--- The comparison between
linear--theory predictions (dashed curves) and results from the
analysis of mock samples for the velocity correlation function
$\psi_1(r)$ (in units of $10^4\vel$). Mock samples are extracted from
an N--body simulation of the $\Lambda 0.4$ model.  Upper and lower
panels refer to uniform weighting and distance--error weighting,
respectively. Error bars are the $1\sigma$ scatter among the set of
216 mock samples.}
\vspace{5mm}

\subsection{Testing the analysis method}
Since the mock samples have been generated by reproducing the
positions of real galaxies, their corresponding ${\cal A}(r)$ is the
same as for the real SFI sample.  For each cosmological model we
compute in linear theory the expected $\psi_1$ (eq.[\ref{eq:ppsi1}])
and compare it to the distribution of values obtained from the mock
samples using eq.(\ref{eq:psi1}). We plot in
Figure 3 the results of this comparison for the
$\Lambda 0.4$ case, for both uniform (upper panel) and distance--error
(lower panel) weightings.  Filled circles represents $\psi_1(r)$ as
estimated by averaging over the set of $N_{mock}=216$ model samples
and the errorbars are the 1$\sigma$ scatter, arising from both cosmic
variance and observational uncertainties.  As a basic result, it turns
out that, for both weighting schemes, linear theory is always adequate
to describe the expected velocity correlation function for samples
having the same selection effects as the SFI, once they are accounted
for by the ${\cal A}(r)$ quantity. Any residual discrepancy on small
($\mincir 1500\vel$) scales, which are probably due to sampling
effects or to residual non--linearities, are well within the 1$\sigma$
scatter. Furthermore, we also remind that since the SFI sample only
contains field spirals, we expect their dynamics to be even closer to
linear theory that that of the N--body particles belonging to the mock
samples, that we did not attempt to select so as to avoid high--density
regions.

We checked the relative contribution to the errors from the cosmic
scatter and from the uncertainties in the peculiar velocity
measurements, using a set of mock samples where peculiar velocities
are not perturbed according to ITF distance errors, so that only the
effect of the cosmic scatter is present.  It turns out that the cosmic
scatter is clearly dominant at $r < 3500\vel$, with the TF scatter
contributing $< 20\%$ and becoming relevant only at larger scales.
The distance--error weighting scheme generates a larger scatter, as a
consequence of the fact that this method amounts to reducing the
effective volume where $\psi_1(r)$ is computed. For this reason, in
the following we will take the uniform weighting as the reference
analysis method to constrain model parameters.


\subsection{Estimating $\psi_1$ uncertainties}
Having demonstrated that linear theory provides reliable predictions
for $\psi_1$, the next information that one needs is the uncertainty
to be associated to such predictions. 
In order to do so, 
we estimate from the set of mock samples the elements of the covariance 
matrix, ${\cal C}^{ij}$, which are defined as
\be
{\cal C}^{ij}\,=\,{1\over
N_{mock}}\,\sum_{l=1}^{N_{mock}}\left(\psi_{1,l}^i-\bar\psi_1^i\right)\,
\left(\psi_{1,l}^j-\bar\psi_1^j\right)\,.
\label{eq:cova}
\ee
Here $\psi_{1,l}^i$ is the value of the velocity correlation function
at the $i$--th separation bin for the $l$--th mock sample, while
$\bar\psi_1^i$ is its average value estimated over the $N_{mock}$
samples. 

Figure 4 shows the comparison between results from the
$\Lambda 0.4$ and EdS models, by plotting the quantities ${\cal
C}^{ij}/\psi_1^i\psi_1^j$.  According to its definition, this quantity
describes the relative covariance of the $\psi_1$ values at different
separations.  The top panels show the results for the diagonal
(variance) terms, while the other panels show the off--diagonal terms,
illustrating different rows in the covariance matrix.  The first thing
to note is the large cross-correlation between the results of the
different bins, which are comparable to the variances, and therefore
cannot be ignored when using the $\psi_1$ statistic to constrain
cosmological models.

In addition, it is apparent from the figure that, apart from small
differences due to statistical fluctuations, the two models have the
same amount of relative covariance.  This is not unexpected since, to
a first approximation, the long--wave perturbations which generate the
cosmic scatter, are also responsible for the $\psi_1$ signal, so as to
make the relative scatter fairly constant.  Noticeable differences
occur only at relatively large separations, $> 3500$ km s$^{-1}$,
where the observational uncertainties become more dominant, thus
increasing the total scatter and suppressing the discriminative power
of $\psi_1(r)$.  For this reason, in the following we will compare
linear--theory predictions and SFI results only for $r\le 3500\vel$,
where the relative uncertainties are essentially the same for the two
considered models. We note that, since $\Lambda 0.4$ and EdS have
rather different values for both $\eta_8$ and for the power--spectrum
shape, we can quite confidently conclude that the relative scatter for
$\psi_1(r)$ is model--independent, at least for the range of models
and scales of interest, while its absolute value is not.

\begin{figure*}[t]
\vspace{-1.5truecm}
\includegraphics{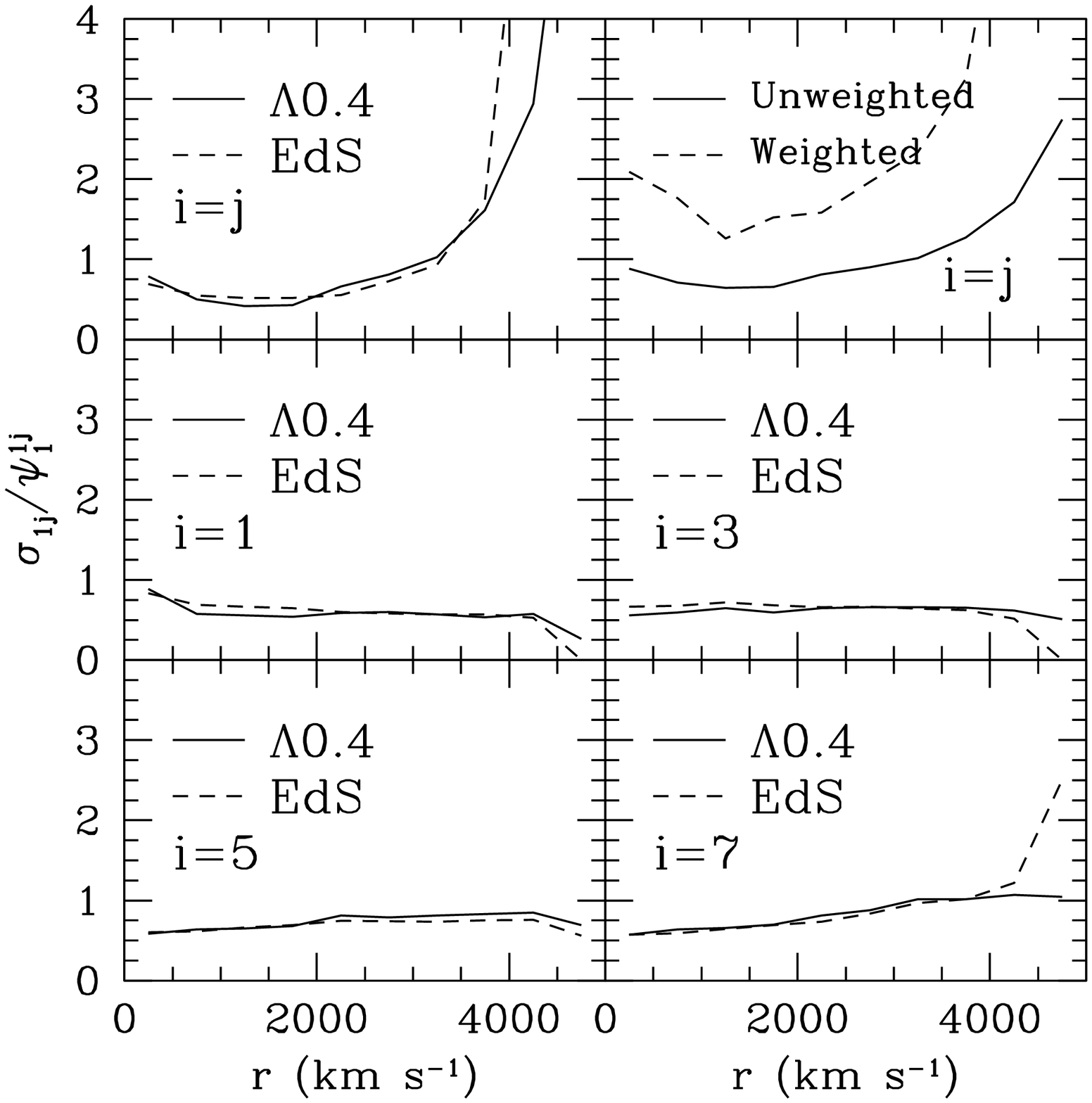}
$\ \ \ \ \ \ $\\
\vspace{15.5truecm}
$\ \ \ $\\
{\small\parindent=3.5mm {Fig.}~4.--- The elements of the relative
covariance matrix, ${\cal C}^{ij}/\psi_1^i\psi_1^j$, for SFI mock
samples extracted from EdS and $\Lambda$0.4 simulations.  The top
panels shows the diagonal (variance) terms, with the top right panel
comparing the variance for unweighted and error--weighted estimates of
$\psi_1(r)$.  The other panels are for the off--diagonal terms and
show different rows in the covariance matrix (see text).
}
\vspace{5mm}
\end{figure*}

In the top right panel of Fig. 4 we compare the diagonal
terms for the $\Lambda 0.4$ mock samples for $\psi_1$ computed
according to uniform and distance--error weighting schemes. It is
apparent that the distance--error weighting is associated with larger
error bars, as was already shown in Figure 3.

Based on these results we, therefore, conclude that: {\em (a)} the
errors of individual $\psi_1$ bins are significantly correlated; {\em
(b)} a general recipe can be devised for the $\psi_1$ uncertainties,
whose relative amount is fairly independent of the cosmological model;
and {\em (c)} that the size of such errors is smaller when $\psi_1$ is
estimated according to the uniform--weighting scheme.

\section{CONSTRAINING COSMOLOGICAL MODELS}
\label{cons}
Based on the results obtained so far, we will now use
eqs.(\ref{eq:ppsi1}) and (\ref{eq:psi}) as a model prediction for
$\psi_1$. As for the model power spectrum, we express it as
$P(k)=A\,k\,T^2(k)$ where we assume a Harrison--Zel'dovich shape 
on large scales. The transfer function, $T(k)$, is taken to be
\ba
& &T(q)\, = \, {{\mbox{\rm ln}}(1+2.34q)\over
2.34q}\times\nonumber \\
& & \left[1+3.89q+(16.1q)^2+(5.46q)^3+(6.71q)^4\right]^{-1/4}\,, 
\label{eq:tk}
\ea
where $q=k/\Gamma h$ and $\Gamma$ is the so--called shape parameter.
For $\Gamma\simeq \Omega_mh$, eq.(\ref{eq:tk}) provides the transfer
function for CDM models with a negligible baryon fraction (Bardeen et
al. 1986). More generally, it can be seen as a phenomenological
expression, with $\Gamma$ a parameter to be fixed by observational
constraints. As for the amplitude of the power spectrum, it is
customary to express it in terms of $\sigma_8$. Following
eqs.(\ref{eq:psi}), the velocity correlation function $\psi_1(r)$ is
then entirely specified in linear theory by the two parameters
$\Gamma$ and $\eta_8$. 

Despite the errorbars being 
so large such that the $\psi_1$ detection is only marginally different
from zero in each individual bin (cf. Figs. 2 and
3), its determination at different scales does allow
to place significant constraints on the $\eta_8$--$\Gamma$ plane.
In order to provide constraints on these
parameters, we compute the weighted $\chi^2$ between the SFI
correlation function, $\psi^{SFI}_1$, and that from model predictions,
$\psi^{mod}_1$:
\ba
\chi^2=\sum_{i,j}&&\!\!\!\!\!\!\!\!\!\!\!
\left[\psi^{SFI}_1(r_i)-\psi^{mod}_1(r_i)\right]\,{\cal
C}_{ij}^{-1}\nonumber \\
&&\!\!\!\!\!\!\!\!\!\!\!
\left[\psi^{SFI}_1(r_i)-\psi^{mod}_1(r_i)\right]\,.
\label{eq:chi2}
\ea 
Here, ${\cal C}_{ij}^{-1}$ are the elements of the inverse of the
covariance matrix, as calibrated from the mock samples, and the sums
are over the radial bins of $500\vel$ width, for separations $r\le
3500$ km s$^{-1}$.  The probability for model rejection is estimated
by assuming a $\chi^2$ statistic, from the value of
$\Delta\chi^2=\chi^2-\chi^2_{min}$, where $\chi^2_{min}$ is the
absolute minimum value.

\includegraphics{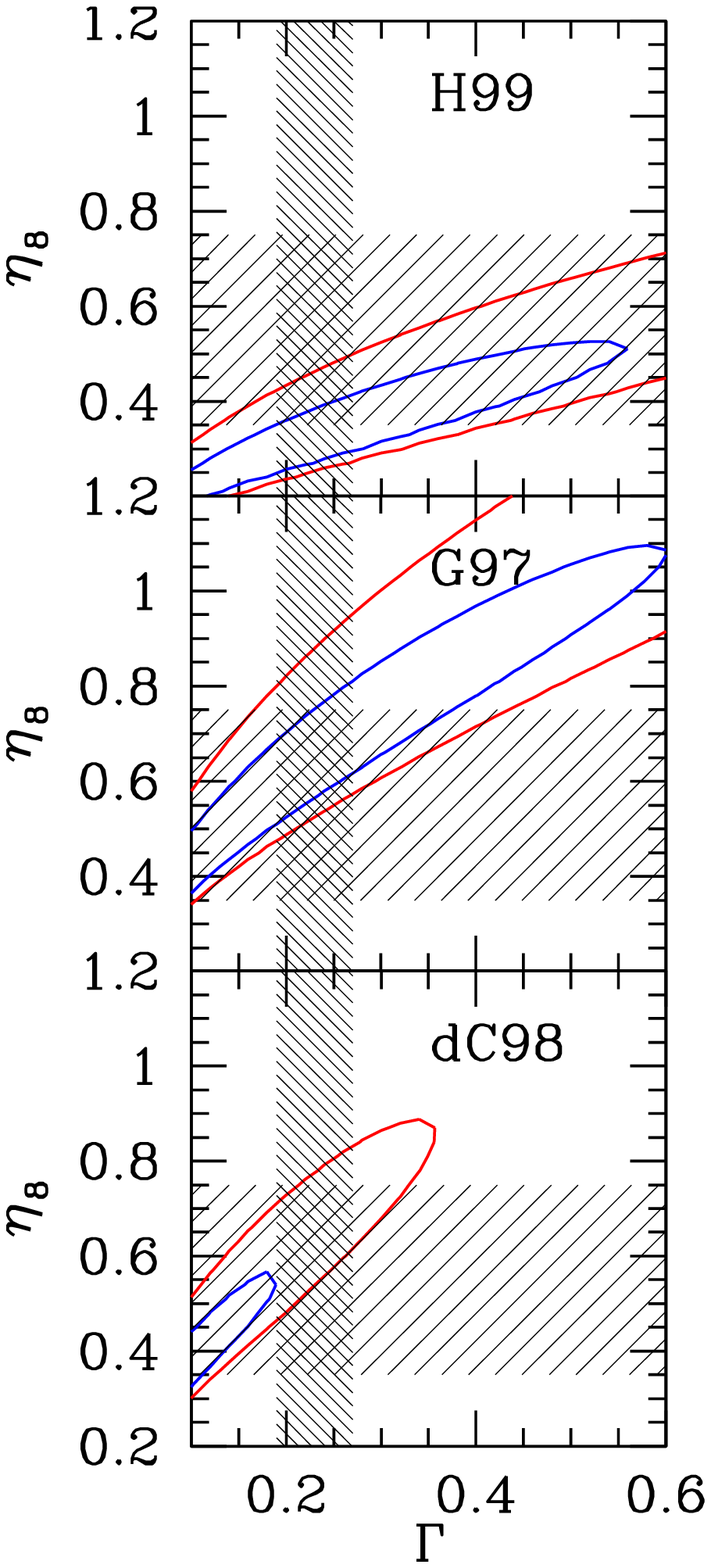}
$\ \ \ \ \ \ $\\
\vspace{15.truecm}
$\ \ \ $\\

\vspace{-1.truecm}
{\small\parindent=3.5mm {Fig.}~5.--- The 1$\sigma$ and $2\sigma$
contours in the $\eta_8$--$\Gamma$ plane from the analysis of the
velocity correlation function, $\psi_1(r)$, for different calibrations
of the inverse Tully--Fisher relation. The horizontal shaded area
corresponds to the $90\%$ confidence level constraints on $\eta_8$
from the analysis of the Giovanelli et al.  (1997a,b) r.m.s. cluster
peculiar velocities (Borgani et al. 1997). The vertical shaded area is
the $95\%$ confidence level constraint on the shape parameter from the
power--spectrum of APM galaxies (Liddle et al. 1996). }
\vspace{2mm}

In Figure 5 we plot the iso--$\Delta\chi^2$ contours for
the three ITF calibrations of the SFI sample that were discussed in
Section 2. Internal and external contours correspond to
$\Delta\chi^2=2.30$ and 6.17, respectively, thus providing the
1$\sigma$ and $2\sigma$ confidence levels for two significant
parameters.  The corresponding minimum values of the $\chi^2$ per
degree of freedom are 1.67, 0.80, and 0.78, for the H99, G97 and dC98
calibrations, respectively.  In all cases, the best-fitting model
seems to provide an acceptable fit.  This value for the H99
calibration is somewhat large, however it corresponds to only $\sim 1
\sigma$ deviation for a $\chi^2$ statistic with five degrees of
freedom.  The fact that such $\chi^2$ values are around unity
indicates that our error model is realistic.

The vertical shaded areas represent the $95\%$ confidence level
interval on the shape parameter, as derived by Liddle et al.  (1996)
from the power--spectrum of APM galaxies. The horizontal shaded areas
represent the $90\%$ confidence level on $\eta_8$ derived by Borgani
et al. (1997) from an analysis of the r.m.s. peculiar velocity of SCI
clusters (Giovanelli et al. 1997a).  All these constraints intersect
our $2\sigma$ confidence regions.

For the H99 and G97 ITF calibrations, the constraints
in the $\eta_8$--$\Gamma$ plane can be cast in the form 
\be 
\eta_8\,=\,\eta_{8,0}\times
\left({\Gamma \over 0.2}\right)^{0.5}\,,
\label{eq:sigga}
\ee 
with $\eta_{8,0}=0.30^{+0.12}_{-0.07}$ and
$\eta_{8,0}=0.58^{+0.22}_{-0.12}$ for the two above calibrations,
respectively (errorbars correspond to $2\sigma$ c.l.). The
asymmetry in the errors is due to the fact that, as $\eta_8$ is
increased from its best--fitting values, larger absolute errors are
assigned to $\psi_1$, since the relative scatter is taken to be
constant (cf. \S4.3).  Thus, larger values of $\eta_8$ tend to be
excluded at
a lower significance than smaller values.  As for the dC98
calibration, the corresponding constraints show a somewhat steeper
$\Gamma$--dependence of $\eta_8$ with values of $\Gamma \mincir 0.35$
ruled out at about $2\sigma$ c.l.  It is interesting to note that, for
$\Gamma\simeq 0.2$, this result agrees with $\beta=0.6\pm0.1$, as
found by da Costa et al. (1998), for an almost unbiased IRAS galaxy
distribution.

We show in Figure 6 the variation of $\Delta\chi^2$
around its minimum as a function of $\eta_8$, in order to show the
effect of changing other assumptions underlying our analysis.  In all
the panels, the solid curve refers to constraints from the H99 ITF
calibration, for a fixed shape-parameter $\Gamma=0.2$ and $\log W >
2.25$ for the line--width of SFI galaxies.  

\begin{figure*}[t]
\vspace{-1.5truecm}
\includegraphics{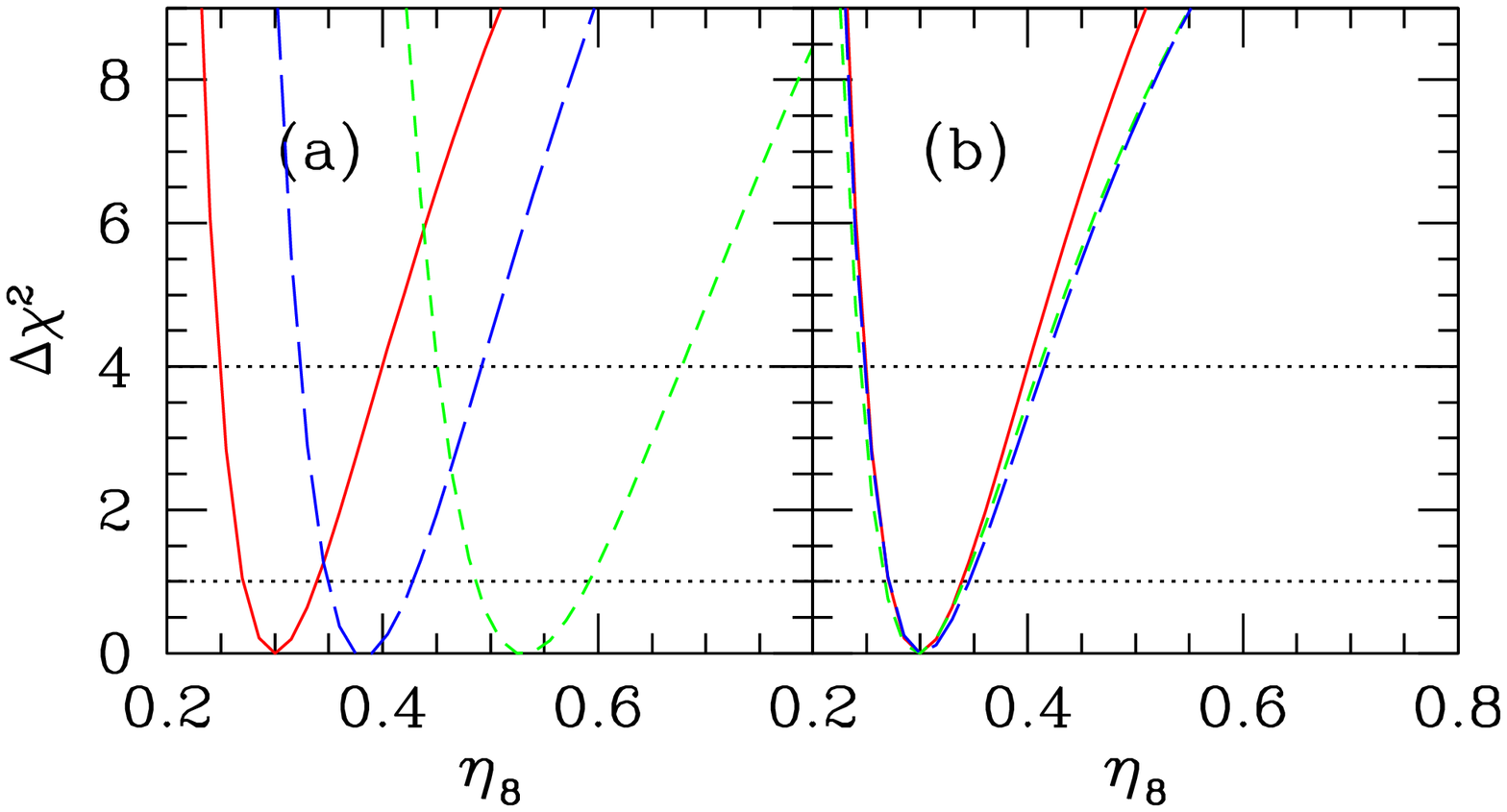}
$\ \ \ \ \ \ $\\
\vspace{8.5truecm} 
$\ \ \ $\\ 

{\small\parindent=3.5mm {Fig.}~6.---
Variation of $\Delta \chi^2$ around its minimum value as a function of
$\eta_8$.  In both panels the solid curve corresponds to $\Gamma=0.2$,
uniform weighting in the estimate of $\psi_1(r)$, ITF calibration by
H99a,b, with the best--fitting value of the zero--point, and $\log W >
2.25$ for the galaxy line--width.  Panel (a): effect of changing the
zero--point of the ITF relation; short-- and long--dashed lines are
for shifting it by 0.1 magnitudes upwards and downwards, respectively.
Panel (b): effect of increasing the limiting line--width; short-- and
long--dashed lines are for $\log W> 2.3$ and $\log W> 2.4$,
respectively.}

\vspace{5mm}
\end{figure*}

As demonstrated already
(Fig. 2, top panel), our results are insensitive to the choice of
galaxy weighting, and we adopt here throughout the uniform
weighting. As is illustrated here [panel $(b)$], changing the limiting
line--width of the sample also has a negligible effect on our results
which are virtually unchanged as we increase it from 2.25 to 2.40.  We
find as well that our constraints do not depend on the specific choice
of binning used in the computation of $\psi_1(r)$.  The effect of the
zero--point uncertainty is shown in panel $(a)$.  As was illustrated
also in the lower panel of Figure 2, the results are quite sensitive
to such changes, and a negative shift of the ITF zero--point by 0.1
mag leads to a sizeable increase of $\eta_8$ from $\simeq 0.3$ to
$\simeq 0.55$.  For higher values of $\Gamma$, this change would
similarly correspond to higher values of $\eta_8$, e.g for
$\Gamma=0.4$, $\eta_8$ would increase from $\simeq 0.4$ to $\simeq
0.8$, and its effect is generally comparable to that of varying the
ITF calibration.

Despite the fact that the constraints on cosmological parameters drawn
from the $\psi_1$ statistics are quite sensitive to the details of the
ITF calibrations, some conclusions can still be drawn.  First, the
constraints on the velocity power--spectrum normalization, $\eta_8$,
depend on the $P(k)$ shape, as a consequence of the fact that we are
probing velocity fields on scales larger that the 8\hm normalization
scale.  Second, assuming $\Gamma \simeq 0.2$, as indicated by galaxy
clustering data, implies power--spectrum amplitudes which can be
different by up to a factor two, but are still generally consistent
with independent observational constraints. For instance, the local
abundance of galaxy clusters to a first approximation also provides a
constraint on $\eta_8=0.5$--0.6 (e.g. Eke, Cole \& Frenk 1996; Girardi
et al. 1998, and references therein).

Our results for $\eta_8$ can also be compared with those obtained by
Zaroubi et al. (1997) and FZ99, who estimated the mass power spectrum
by a maximum likelihood (ML) analysis of the peculiar velocities of
the Mark III and the SFI samples, respectively.  These estimates are
then translated to constraints on $\eta_8$ by integrating over the
corresponding spectra. Both works consistently found $\eta_8 \simeq
0.8\pm 0.2$ at $90\%$ c.l. and a preferred value of $\Gamma\simeq
0.4\pm 0.2$. As the application of the ML analysis for the SFI sample
has been performed using the G97 calibration, it is most suitable to
compare the FZ99 results with those reported in the central panel of
Fig. 5. It turns out that the confidence regions coming
from the ML and $\psi_1$ analyses do overlap over a significant
portion of the $\eta_8$--$\Gamma$ plane. For $\Gamma=0.4$, the
$\psi_1$ analysis gives $\eta_8=0.85^{+0.17}_{-0.10}$.  The main
difference being the dependence of the $\eta_8$ constrains on $\Gamma$
in the $\psi_1$ analysis, such that for lower values of $\Gamma \simeq
0.2$ the preferred $\eta_8$ values are somewhat smaller than those
obtained in the ML analysis.

One should also bear in mind the different sensitivities of these two
analyses. As demonstrated in Figs. 5 and 6,
the $\psi_1$ analysis is sensitive to the ITF calibration, while it is
robust to changing the limiting line--width. On the other hand, the ML
analysis is remarkably robust to changes in TF calibration (e.g.
Fig. 8 in FZ99), while it is more sensitive to the pruning of SFI
galaxies at different line--widths.  For these reasons, these two
methods should be regarded as complementary and both worth to be
applied to a given data set.

\section{CONCLUSION} 
\label{conc}
In this paper we have presented an analysis of the velocity
correlation function, $\psi_1(r)$, for the SFI sample of Sbc--Sc
galaxy peculiar velocities based on the infrared TF distance indicator
calibrated using a sample of cluster galaxies (Giovanelli et
al. 1997a,b; Haynes et al. 1999a,b). In order to minimize
uncertainties related to Malmquist bias corrections, we performed the
analysis using the redshift--space positions of galaxies and the ITF
distance indicator. Three different ITF calibrations for the SFI
sample have been examined in our analysis: one based on an updated
version of the SFI sample presented by Haynes et al. (1999a,b, H99),
that presented by Giovanelli et al. (1997b, G97) and that obtained by
da Costa et al. (1998, dC98).

The final goal of our analysis is to place constraints on the
amplitude and the shape of the fluctuation power--spectrum, by
comparing $\psi_1(r)$ from SFI and from linear--theory predictions of
cosmological models.  For this purpose, we needed to verify the
reliability of linear--theory to predict $\psi_1(r)$ for a sample
having the same galaxy positions and observational uncertainties as
the SFI one, and to estimate the associated uncertainties due to
cosmic scatter and observational uncertainties.  These two goals have
been achieved by comparing linear--theory predictions to results from
the analysis of a large set of mock SFI samples, extracted from
N--body simulations.

We have found that linear--theory provides a rather accurate
description of the $\psi_1(r)$ estimated from the mock samples, over
the whole scale range considered ($r\le 5000\vel$; cf. Figure
3). This confirms that both sparse sampling effects
and residual non--linearities have a minor impact on our analysis.  We
have also shown that the relative covariance in $\psi_1(r)$ among the
set of mock samples is roughly independent of the cosmological models,
thus allowing for a simple treatment of the associated errors.

In general, we find that our analysis constrains a degenerate ridge in
the $\eta_8$--$\Gamma$ plane. For the H99 and G97 ITF calibrations, we
find $\eta_8=\eta_{8,0} (\Gamma/0.2)^{0.5}$, with
$\eta_{8,0}=0.30^{+0.12}_{-0.07}$ and
$\eta_{8,0}=0.58^{+0.22}_{-0.12}$ for the two above calibrations,
respectively, at the $2\sigma$ level (cf. Figure 5).  The
dC98 exhibits a stronger tendency for lower values of the shape
parameter, constraining $\Gamma\mincir 0.35$ at the $2\sigma$ level,
and is consistent with the G97 calibration in that range.  These
constraints are robust to variations of the galaxy weighting scheme
(cf. Figure 2) and to changes in the choice of the
limiting galaxy line--width (cf. Figure 6), but are,
clearly, very sensitive to uncertainties in the calibration
details, such as the zero-point of the TF relation.

In any case, the results presented here indicate that the large--scale
velocity field can be brought into agreement with the low fluctuation
amplitude implied at $\sim 10$\hm scale by the abundance of galaxy
clusters (e.g. Eke et al. 1996, Girardi et al. 1998), for
power--spectrum shapes which are consistent with large--scale
clustering data (e.g. Liddle et al. 1996), while higher amplitudes are
allowed for larger values of the shape parameter.  Our constraints on
the $\eta_8$--$\Gamma$ plane for the ITF G97 calibration and those
from the maximum--likelihood (ML) analysis for the G97 direct TF
relation by Freudling et al. (1999, FZ99) are quite consistent for
$\Gamma\magcir 0.3$.  Since the ML and the $\psi_1$ methods are
sensitive to different degrees to different aspects of the analysis
(i.e., TF calibration and limiting line--width), they should be
regarded as complementary approaches for extracting cosmological
constraints from large--scale cosmic flows.

\acknowledgments We acknowledge useful discussions with Adi Nusser,
Avishai Dekel and Saleem Zaroubi. We thank the referee Michael Strauss
for many useful comments, which improved the presentation of the
results.  We thank Hugh Couchman for the generous sharing of his
adaptive $P^3M$ code. We are grateful to the ESO Visitors fund for
supporting visits to Garching by SB, RG, MPH, and IZ. SB acknowledges
ICTP and SISSA in Trieste, for the hospitality during several phases
of preparation of this work.  IZ was supported by the DOE and the NASA
grant NAG 5-7092 at Fermilab.

\end{multicols}

\end{document}